\documentclass[11pt]{article}

\usepackage{graphicx}

\usepackage{multicol} 

\usepackage{color}

\definecolor{rosso}{cmyk}{0,1,1,0.4}
\definecolor{rossos}{cmyk}{0,1,1,0.55}
\definecolor{rossoc}{cmyk}{0,0.5,1,0.2}
\definecolor{blu}{cmyk}{1,1,0,0.3}
\definecolor{blus}{cmyk}{1,1,0,0.6}
\definecolor{blucc}{cmyk}{1,0.4,0.2,0}
\definecolor{viola}{cmyk}{0,1,0,0.6}
\definecolor{viola2}{cmyk}{0,1,0.2,0.6}
\definecolor{verde}{cmyk}{0.92,0,0.59,0.25}
\definecolor{verdec}{cmyk}{0.92,0,0.59,0.15}
\definecolor{verdes}{cmyk}{0.92,0,0.59,0.4}
\font\tenrsfs=rsfs10 at 12pt
\font\sevenrsfs=rsfs7
\font\fiversfs=rsfs5
\newfam\rsfsfam

\textfont\rsfsfam=\tenrsfs
\scriptfont\rsfsfam=\sevenrsfs
\scriptscriptfont\rsfsfam=\fiversfs
\def\mathscr#1{{\fam\rsfsfam\relax#1}}
\def\Lag{\mathscr{L}}

\newcommand{\fig}[1]{~\ref{fig:#1}}
\newcommand{\eq}[1]{~{\rm (\ref{eq:#1})}}

\newcommand{\GeV}{\,{\rm GeV}}
\newcommand{\TeV}{\,{\rm TeV}}
\def\circa#1{\,\raise.3ex\hbox{$#1$\kern-.75em\lower1ex\hbox{$\sim$}}\,}

\newcommand{\bW}{{\bar W}}
\newcommand{\bB}{{\bar B}}

\newcommand{\beq}{\begin{equation}}
\newcommand{\eeq}{\end{equation}}
\newcommand{\bea}{\begin{eqnarray}}
\newcommand{\eea}{\end{eqnarray}}

\def\circa#1{\,\raise.3ex\hbox{$#1$\kern-.75em\lower1ex\hbox{$\sim$}}\,}
\makeatletter

\def\art{\@ifnextchar[{\eart}{\oart}}
\def\eart[#1]#2#3#4#5#6{{\rm #2}, {\em #3 \rm #4} {\rm (#6) #5} ({#1})}
\def\hepart[#1]#2{{\rm #2, #1}}
\newcommand{\oart}[5]{{\rm #1}, {\em #2 \rm #3} {\rm (#5) #4}}

\newcounter{alphaequation}[equation]
\def\thealphaequation{\theequation\hbox to
0.6em{\hfil\alph{alphaequation}\hfil}}
\def\eqnsystem#1{
\def\@eqnnum{{\rm (\thealphaequation)}}
\def\@@eqncr{\let\@tempa\relax \ifcase\@eqcnt \def\@tempa{& & &} \or
  \def\@tempa{& &}\or \def\@tempa{&}\fi\@tempa
  \if@eqnsw\@eqnnum\refstepcounter{alphaequation}\fi
\global\@eqnswtrue\global\@eqcnt=0\cr}
\refstepcounter{equation} \let\@currentlabel\theequation \def\@tempb{#1}
\ifx\@tempb\empty\else\label{#1}\fi
\refstepcounter{alphaequation}
\let\@currentlabel\thealphaequation
\global\@eqnswtrue\global\@eqcnt=0 \tabskip\@centering\let\\=\@eqncr
$$\halign to \displaywidth\bgroup \@eqnsel\hskip\@centering
$\displaystyle\tabskip\z@{##}$&\global\@eqcnt\@ne
\hskip2\arraycolsep\hfil${##}$\hfil& \global\@eqcnt\tw@\hskip2\arraycolsep
$\displaystyle\tabskip\z@{##}$\hfil
\tabskip\@centering&\llap{##}\tabskip\z@\cr}
\def\endeqnsystem{\@@eqncr\egroup$$\global\@ignoretrue} \makeatother

\newcommand{\sW}{s_{\rm W}}
\newcommand{\cW}{c_{\rm W}}

\begin{document}

\thispagestyle{empty}

\begin{flushright}
{CERN-PH-TH/2004-075\\
IFUP--TH/2004--13\\
hep-ph/0405040\\
UAB-FT-565
}
\end{flushright}
\vspace{1cm}

\begin{center}
{\LARGE \bf \color{rossos}
Electroweak symmetry breaking\\[2mm]
 after LEP1 and LEP2
}\\[1cm]

{
{\large\bf Riccardo Barbieri}$^a$,
{\large\bf Alex Pomarol}$^b$,\\[2mm]
  {\large\bf Riccardo Rattazzi}$^{c}$\footnote{{On leave of absence from INFN, Pisa, Italy.}}, {\large\bf Alessandro Strumia}$^{d}$
}  
\\[7mm]
{\it $^a$ Scuola Normale Superiore, Pisa, and INFN, Italia } \\[3mm]
{\it $^b$ IFAE, Universitat Aut{\`o}noma de Barcelona,
08193 Bellaterra (Barcelona), Spain }\\[3mm]
{\it $^c$ Physics Department, Theory Division, CERN, CH-1211 Geneva 23, Switzerland}\\[3mm]
{\it $^d$ Dipartimento di Fisica dell'Universit{\`a} di Pisa
and INFN, 
Italia}\\[1cm]
\vspace{1cm}
{\large\bf\color{blus} Abstract}
\end{center}
\begin{quote}
{\large\noindent\color{blus}
In a generic ``universal'' theory of electroweak symmetry breaking, simple symmetry considerations and absence of
 tuning imply that
 heavy new physics affects the low-energy data through four parameters. These include and properly extend the generally insufficient  $S$ and $T$. 
Only by adding the LEP2 data to the global electroweak fit,  can all these four form factors 
be determined and deviations from the SM be strongly constrained. Several of the recently proposed models (little Higgs,  gauge bosons in extra dimensions or Higgsless models in 5D) are recognized to be ``universal'' in a straightforward way  after a proper definition of the effective vector boson fields. Among various applications, we show that proposed Higgsless models in 5D, when calculable, do not provide a viable description of
 electroweak symmetry breaking  in their full range of parameters.
}
\end{quote}


\newpage

\setcounter{page}{1}
\setcounter{footnote}{0}

\section{Introduction and statement of the problem}

The physical mechanism underlying Electroweak Symmetry Breaking (EWSB) remains unknown.
Its description in the Standard Model (SM) is not fully satisfactory, with reasons that motivate a modification of the SM at energies close to the Fermi scale. Examples of recent theoretical attempts along these directions include little Higgs models \cite{littleH} and models in 5D with or without a 
Higgs \cite{Csaki:2003dt, BPR}.

While waiting for the LHC to provide a thorough experimental exploration of the energy scales relevant to  EWSB, we find it useful to reconsider the problem of describing the 
phenomenology of EWSB in a rather model independent way. There is one main reason for doing this.  In the analysis of some models, as we are going to see, the traditional use of 3 parameters, 
$S$, $T$ and $U$ \cite{tech,pt,alba}  is determined more by the limited  information provided by the measurements around the $Z$-pole rather than by a satisfactory theoretical background. 
It is therefore important  that this information can now be complemented by the one available from LEP2, which requires a suitable extension of the  standard analysis. The comparison of the models mentioned above with current experimental constraints, where the use of the traditional parameters may also be a source of conceptual confusion, provides  clear examples for the usefulness of this extension.

As physically motivated and customary, we shall consider  ``universal'' theories, where the deviations from the SM reside only in the self-energies of the vector
 bosons.  Moreover we want to focus on the case in which these deviations are associated with new physics at an energy scale sensibly higher than 
the LEP2 energy.  Then it is useful to split the exact vacuum polarizations as the sum of two pieces.
The first is a local tree level term, while the second  is purely due to SM loops (this second term
is also non-analytic due to the presence of light fermions). 
In an effective Lagrangian approach, the effects of new physics can then be fully parametrized by  the first term, corresponding to the
tree level 
transverse vacuum polarization amplitudes $ \Pi_V(q^2)$ where $V=\{W^+W^-, W_3 W_3, BB,W_3 B\}$. These amplitudes, according to our assumptions, can be expanded 
in $q^2$
\begin{equation}\label{eq:espansione}
 \Pi_V(q^2) \simeq  \Pi_V(0) + q^2  \Pi'_V(0) +\frac{(q^2)^2}{2!} \Pi''_V(0)  +\cdots\, .
 \end{equation}
 It is important to realize that the category of ``universal'' models is broader than often thought. In particular it includes the possibility that new heavy vector
 states exist, as long as they are coupled to the SM fermions via the usual 
 ${\rm SU}(2)_L\otimes {\rm U}(1)_Y$ currents. This just means than the only gauge interaction of the light fermions (apart from QCD) is
\begin{equation}
\Lag_{\rm int}=\bar\Psi \gamma^\mu \left (T^a{\bar W}^a_\mu+ Y {\bar B}_\mu \right )\Psi\, ,
\label{fermions}
\end{equation}
though ${\bar W}^a$ and $\bar B$  do not coincide in general with the ``light'' vector bosons of the SM. Instead they are 
a mixture of the light with new heavy vector bosons. The self-energies we refer to in eq.~(\ref{eq:espansione})     are therefore the  self-energies of these interpolating fields, as they are defined by the very eq.~(\ref{fermions}) including their normalization. This will be further illustrated in section~\ref{examples}.

As we shall explain below, in a wide class of models satisfying some reasonable requirements, 
it is necessary and sufficient, for a consistent analysis of  the electroweak data,
to consider the expansion in eq.\eq{espansione} up to ${\cal O}(q^4)$.  At this order,
 given the four self-energies, there is naively a total of 12 coefficients. 
Three of them, however, are absorbed in the definition of 
\begin{equation}\label{eq:norm}
\frac{1}{g^2} =  \Pi'_{W^+W^-}(0),\qquad
\frac{1}{g^{\prime 2}} =  \Pi'_{BB}(0),\qquad
v^2 = -2  \Pi_{W^+W^-}(0)\approx (174\GeV)^2\, .
\end{equation}
(notice that we find convenient to choose a non canonical normalization of
the vector bosons).
Furthermore, requiring the masslessness of the photon, coupled to $Q=T_3+Y$, 
implies two relations among the zeroth order coefficients $\Pi_V(0)$. 
Altogether this leaves 7 undetermined parameters, $\widehat{S}, \widehat{T}, \widehat{U}, V, X, Y, W$, defined in 
Table~\ref{tab:STUVXYW}. 
The notation for the 3 residual coefficients up to order $q^2$ makes clear reference to the traditional ones, $S, T, U$~\cite{pt}: the actual relation is  $S= 4s_{\rm W}^2  \widehat{S}/\alpha\approx 119\, \widehat{S}$,
  $T= \widehat{T}/\alpha\approx 129\, \widehat{T}$,
  $U=-4 s_{\rm W}^2 \widehat{ U}/\alpha$. As a natural extension of this formalism, Table~\ref{tab:STUVXYW} also includes an additional form factor in the
 QCD sector, which is  not related to EWSB and which we will henceforth neglect.

\begin{table}
$$\hspace{-4mm}
\begin{array}{rclrlcc}
\multicolumn{3}{c}{\hbox{Adimensional form factors}}&
\multicolumn{2}{c}{\hbox{operators}}& \hbox{custodial} & \hbox{SU(2)$_L$}\\ \hline
g^{-2}{\color{blus}\widehat{S}} &=& \Pi'_{W_3 B}(0) & {\cal O}_{WB}~=&(H^\dagger \tau^a H) W^a_{\mu\nu} B_{\mu\nu} /gg'\!\!\!&+&-\\[1mm]
g^{-2}M_W^2{\color{blus} \widehat{T} }&=& \Pi_{W_3 W_3}(0)-\Pi_{W^+W^-}(0)\!\!\!
& {\cal O}_H~=&|H^\dagger D_\mu H|^2&-&-\\[1mm]
-g^{-2}{\color{blus} \widehat{U}} &=&\Pi'_{W_3 W_3}(0)-\Pi'_{W^+W^-}(0)\!\!\! &-&&-&-\\[1mm]
2g^{-2}M_W^{-2}{\color{blus} V} &=& \Pi''_{W_3 W_3}(0) -\Pi''_{W^+W^-}(0)\!\!\! & -&&-&-\\[1mm]
2g^{-1} g^{\prime -1}M_W^{-2}{\color{blus} X}&=&\Pi''_{W_3 B}(0) & -&&+&-\\[1mm]
2g^{\prime-2}M_W^{-2}{\color{blus} Y} &=&\Pi''_{BB}(0) &{\cal O}_{BB}~=&(\partial_\rho B_{\mu\nu})^2/2g^{\prime 2}&+&+\\[1mm]
2g^{-2} M_W^{-2}{\color{blus} W} &=& \Pi''_{W_3 W_3}(0) & {\cal O}_{WW} ~=&(D_\rho W^a_{\mu\nu})^2/2g^2&+&+\\[1mm]
2g_{\rm s}^{-2}M_W^{-2}{\color{blus} Z} &=& \Pi''_{GG}(0) & {\cal O}_{GG} ~=&(D_\rho G^A_{\mu\nu})^2/2g_{\rm s}^2&+&+
\end{array}$$
  \caption{\label{tab:STUVXYW}\em 
  The first column defines the adimensional form factors.
The second column defines the {\rm SU(2)$_L$}-invariant universal dimension-6 
operators,
  which contribute to the form-factors on the same row.
  We use non canonically normalized fields and $\Pi$, see eq.\eq{norm}.
  The $\widehat{S}$, $\widehat{T}$, $\widehat{U}$ are related to the
  usual $S,T,U$ parameters~\cite{pt} as:
    $S= 4s_{\rm W}^2  \widehat{S}/\alpha\approx 119\, \widehat{S}$,
  $T= \widehat{T}/\alpha\approx 129\, \widehat{T}$,
  $U=-4 s_{\rm W}^2 \widehat{ U}/\alpha$. 
  The last row defines one additional form-factor in the QCD sector.
  }
\end{table}

\smallskip
As we shall now explain, the subset $\widehat S, \widehat T, Y, W$ represents the most general parametrization
of new physics effects in Electroweak Precision Tests (EWPT).
Notice that we can group the various 
form factors in 3 different classes according to their symmetry properties. The first class 
is given by $\widehat T$, $\widehat U$ and $V$ as they have the same custodial and
weak isospin breaking quantum numbers. The second class is given by 
$\widehat S$ and $X$, which
 are  custodially symmetric but weak isospin breaking (and odd under the spurionic symmetry  which reverses the sign of 
$B_\mu$ and of the hypercharges of matter fields). Finally $W$ and $Y$, which preserve both custodial and weak isospin, make up the third class.
By going  to ${\cal O}(q^6)$ and higher there would arise no new class but only higher derivative terms in each of the above 3 classes. It is reasonable 
to expect that
coefficients with the same symmetry properties will be related to each other up to trivial factors associated to the number of derivatives: in a model where 
the new physics  comes in at a scale $\Lambda$ we expect  $\widehat U\sim (M_W/\Lambda)^2 \widehat T$, 
$V\sim (M_W/\Lambda)^4 \widehat T$.
Similarly we expect $X\sim (M_W/\Lambda)^2 \widehat S$. 
On the other hand, $W$ and $Y$ are the lowest in their class.\footnote{The leading term in their class is truly
represented by the SM gauge kinetic coefficients $1/g^2$ and $1/g^{\prime 2}$.} As soon as the gap between $M_W$ and $\Lambda$ is big enough, it should be reasonable 
to retain only the lowest derivative term in each class: $\widehat S$, $\widehat T$ , $W$ and $Y$. Neglecting $\widehat U, V, X$ when they are parametrically suppressed
 also makes sense because the experimental sensitivity on them is not higher than for the other four.
Of course one can imagine fine-tuned situations where this reasoning fails. 
On the contrary, although $\widehat S$, $\widehat T$
and $W$, $Y$ have a different number of derivatives
 there is no deep physical reason, in general,
 to expect $\widehat T$ to be
bigger than $\widehat S$ and in turn $\widehat S$ to be bigger than $W,Y$. 
Indeed there are several explicit models where
these 4 quantities give comparable effects. Basically we can associate $\widehat S$ and $\widehat T$ to new physics in the
electroweak breaking sector (both effects break weak isospin), which is the case of technicolor.
On the other hand $W$ and $Y$ are  associated to new structure in the vector channels,
like for instance  vector compositeness or new gauge bosons.
To conclude, we stress, as is made evident from our discussion, that no additional relevant effects are expected by considering terms
with more than 4 powers of momentum.

Our conclusions are not entirely new. The same line of reasoning, applied to ordinary technicolor models, rightly selects 
just $\widehat S$ and $\widehat T$ as relevant parameters~\cite{tech}.  
In addition, keeping the light Higgs field and parametrizing 
new physics effects by higher dimensional operators, one
finds that the leading effects, associated to dimension 6 operators~\cite{Grinstein},\footnote{In~\cite{NRO}
 a complete list  
 of the dimension-6 operators 
affecting precision electroweak data is given. 
In the same list only two of the four operators in Table~\ref{tab:STUVXYW} 
are present.
${\cal O}_{BB}$ and ${\cal O}_{WW}$ are not included.
As shown in~\cite{KK}, these operators are equivalent to proper combinations 
of the operators involving fermions and appearing in the list of~\cite{NRO}.
Names and normalizations of the operators used here agree with~\cite{NRO,KK},
after taking into account that they are here written in terms of non-canonically normalized
gauge bosons.
Imposing supersymmetry does not reduce the number of independent ``universal''
dimension-6 operators~\cite{MaPa}.}
\begin{equation}
\mathscr{L}=\mathscr{L}_{\rm SM}+ 
\frac{1}{v^2}\bigg[  c_{WB} 
 {\cal O}_{WB}+
 c_{H} {\cal O}_{H}+c_{WW} {\cal O}_{WW}+
 c_{BB} {\cal O}_{BB}\bigg]\, ,
\label{eq:NRO}
\end{equation}
correspond precisely to $\widehat S$, $\widehat T$ , $W$ and $Y$:
\begin{equation}
\widehat{S}=2\frac{\cW}{\sW}c_{WB}\ ,\qquad
\widehat{T}=-c_{H}\ , \qquad
W = -g^2 c_{WW}\ ,\qquad 
Y = - g^2c_{BB}\, .
\end{equation}
However we find it useful to emphasize that this parametrization is general. Indeed  
our simple reasoning did not require the 
presence of a Higgs field in the low energy effective theory. Note in particular that
we did not require $\langle H\rangle /\Lambda$ to be a small parameter of our expansion. 

\medskip

EWPT listed in Table~\ref{tab:data}
(and measured mainly at the $Z$-peak by LEP1 experiments,
 but also including the $W$ and top masses and other  measurements)
correspond to 3 ``universal'' observables only, usually named $\varepsilon_1,\varepsilon_2,\varepsilon_3$~\cite{alba},
 and therefore cannot fix the 4 (or more)  form factors possibly generated by  
``universal'' new physics. 
We will show that LEP2 data give 3 additional independent observables, 
here named $\varepsilon_{ZZ},\varepsilon_{Z\gamma},\varepsilon_{\gamma\gamma}$,
that constrain mostly $Y,W$ (or $X,Y,W$, if $X$ is included) as strongly as EWPT.
A combined  analysis is thus needed to properly bound ``universal'' 
new physics scenarios.
These include a  subset of extra dimensional models,
 little Higgs models or Higgsless theories.

The rest of the paper is organized as follows. 
In section~\ref{LEP1} we  express the dependence of the physical 
observables at the $Z$-pole  on  the
 coefficients of Table~\ref{tab:STUVXYW} 
and we summarize the experimental constraints. We also give there the
dependence of the low-energy precision data on the coefficients of Table~\ref{tab:STUVXYW} .
Similarly in section~\ref{LEP2} we consider the information available from LEP2. 
In section~\ref{fit} 
we show the global constraints 
 on the 4 parameters
$\widehat{S}, \widehat{T}, Y, W$ including both EWPT and LEP2.
In section~\ref{examples} we present examples of ``universal''theories
and calculate their  predictions
for 
$\widehat{S}, \widehat{T}, Y, W$.

\begin{table}[t]
$$\begin{array}{rclrl}
\Gamma_Z &=& (2.4952 \pm 0.0023)\GeV  & -0.3\hbox{-}\sigma & \hbox{total $Z$ width} \\
\sigma_h &=&(41.540 \pm 0.037)\hbox{nb}&  1.6\hbox{-}\sigma & \hbox{$e\bar{e}$ hadronic cross section at $Z$ peak}\\
R_h &=& 20.767 \pm 0.025           &  1.1\hbox{-}\sigma & \hbox{$\Gamma(Z\to \hbox{hadrons})/\Gamma(Z\to\mu^+\mu^-)$}\\
R_b &=& 0.21644 \pm 0.00065           &  1.1\hbox{-}\sigma & \hbox{$\Gamma(Z\to b \bar b)/\Gamma(Z\to \hbox{hadrons})$}\\
R_c &=& 0.1718 \pm 0.0031             & -0.2\hbox{-}\sigma & \hbox{$\Gamma(Z\to c \bar c)/\Gamma(Z\to \hbox{hadrons})$}\\
A_{P}^{\tau } &=& 0.1465 \pm  0.0032 &  -0.4\hbox{-}\sigma & \hbox{$\tau$ polarization asymmetry}\\
A_{LR}^e &=& 0.1513 \pm 0.0021        &  1.7\hbox{-}\sigma & \hbox{Left/Right asymmetry in $e\bar{e}$}\\
A_{LR}^b &=& 0.922 \pm 0.02           & -0.6\hbox{-}\sigma &
\hbox{LR Forward/Backward asymmetry in $e\bar{e}\to b\bar{b}$}\\
A_{LR}^c &=& 0.670 \pm 0.026           &  0.1\hbox{-}\sigma & \hbox{LR
 FB asymmetry in $e\bar{e}\to c\bar{c}$}\\
A_{FB}^\ell &=& 0.01714 \pm 0.00095    &  0.8\hbox{-}\sigma & \hbox{Forward/Backward asymmetry in $e\bar{e}\to \ell\bar{\ell}$}\\
A_{FB}^b &=& 0.099 \pm 0.0017         & -2.4\hbox{-}\sigma & \hbox{Forward/Backward asymmetry in $e\bar{e}\to b\bar{b}$}\\
A_{FB}^c &=& 0.067 \pm 0.0026        & 0.1\hbox{-}\sigma & \hbox{Forward/Backward asymmetry in $e\bar{e}\to c\bar{c}$}\\
M_Z &=& 91.1875 \GeV&                                       &\hbox{pole $Z$ mass}  \\

\hline

m_h &>& 114\GeV &&\hbox{Higgs mass}\\ 

G_{\rm F} &=& 1.16637~10^{-5}/\GeV^2&                      & \hbox{Fermi constant for $\mu$ decay}\\

m_t  &=& (178.0\pm4.3)\GeV &      0.3\hbox{-}\sigma                         &\hbox{pole top mass}\\
M_W &=& (80.426 \pm 0.034)\GeV        &  1.1\hbox{-}\sigma & \hbox{pole $W$ mass}  \\ 
\alpha_{\rm s}(M_Z)  &=& 0.118\pm0.003 &  0.0\hbox{-}\sigma                       &\hbox{strong coupling}\\
\alpha_{\rm em}^{-1}(M_Z)  &=& 128.949\pm0.046 &      0.0\hbox{-}\sigma      & \hbox{electromagnetic coupling}\\

\end{array}$$
\caption{\em The  high-energy precision data included in our fit~\cite{LEPEWWG}.
The second column indicates the discrepancy with respect to the best SM fit.\label{tab:data}}
$$\begin{array}{rclrl}
g_L^2 &=& 0.3005 \pm 0.0014     \qquad  \quad    & -3.0\hbox{-}\sigma & \hbox{$\nu_\mu$/nucleon scattering}\\
g_R^2 &=& 0.0310 \pm 0.0011            &  0.5\hbox{-}\sigma & \hbox{$\nu_\mu$/nucleon scattering}\\
Q_W &=& -72.83 \pm 0.49               &  0.1\hbox{-}\sigma & \hbox{atomic parity violation in Cs}\\
A_{\rm PV} &=& (-160\pm 27)\,10^{-9}&  0.8\hbox{-}\sigma & \hbox{M\o{}ller scattering
at $Q^2=0.026\GeV^2$}\qquad\\
\end{array}$$
\caption{\em The  low-energy precision data~\cite{LEPEWWG}. 
We do not include $\nu_\mu$/nucleon scattering data in our global fit.
\label{tab:datalow}}
\end{table}


\section{Electroweak precision observables before LEP2}\label{LEP1}

As mentioned, the effect of ``universal'' theories of EWSB on the 
EWPT listed in table~\ref{tab:data} can be
encapsulated in 3 dimensionless quantities. Here we stick to the parameters $\varepsilon_1, \varepsilon_2, \varepsilon_3$, as defined in~\cite{alba}, which are linearly related to the various observables by universal coefficients only dependent on $\alpha_{\rm s}(M_Z)$ and $\alpha (M_Z)$. 
The $\varepsilon$'s are defined in such a way as to account also for the electroweak radiative correction effects. As such, they are not vanishing even in absence of any deviation from  the SM.

From the dependence of the $\varepsilon$'s on the vacuum polarization amplitudes of the vector bosons~\cite{Barbieri:1991qp}, 
it is immediate to express their dependence on the parameters of Table~\ref{tab:STUVXYW} as 
\begin{eqnsystem}{sys:eps123}
\label{eps123f}
 \varepsilon_1 &=&(+6.0-0.86 \ln\frac{m_h}{M_Z})10^{-3}+ \widehat{T }-W + 2X \frac{\sW}{\cW} - Y \frac{\sW^2}{\cW^2}\, ,\\
\varepsilon_2 &=&(-7.5+0.17 \ln\frac{m_h}{M_Z})10^{-3} +\widehat{U } -W+2X\frac{\sW}{\cW}-V\, ,\\
\varepsilon_3 &=& (+5.2+0.54 \ln\frac{m_h}{M_Z})10^{-3}+\widehat{S} -W+ \frac{X}{\sW\cW}-Y\, .
\label{eps123l}
\end{eqnsystem}
For every $\varepsilon_i$ these equations contain an effective and sufficiently accurate numerical expression for the pure SM contribution.
Our fit takes into account the dependence on $m_t,\alpha_3, \alpha_{\rm em}$.
However in the above equations  we have taken
$m_t = 178\GeV$, $\alpha_3(M_Z) = 0.119$, $\alpha_{\rm em}(M_Z) = 1/128.88$ and we
exhibit only the dependence on the Higgs mass  $m_h$.
In models without a Higgs, the Higgs mass  in the above equations should be interpreted 
as an ultraviolet cutoff of the SM loops provided by the model itself.\footnote{More technically in such theories one should substitute $m_h$ 
with the renormalization scale $\mu$. The resulting explicit $\mu$ dependence of the physical  $\varepsilon_{1,2,3}$ is canceled by
the implicit $\mu$ dependence of the form factors. Notice however that the coefficients  of  $\ln m_h$ in the numerical
approximation of eqs.~(\ref{eps123f})-(\ref{eps123l}) do not exactly correspond
to the analytic one-loop result.}
These terms correspond to infrared logarithms
in the low energy Higgsless theory. 

\medskip

There are 3 experimental parameters $\varepsilon_{1,2,3}$ because
this is all that EWPT
can measure of new physics effects within ``universal'' models. As already mentioned, in some relevant cases the measurement of 
 $\varepsilon_{1,2,3}$  is used to place bounds on the new physics form 
factors $\widehat{S}, \widehat{T}, \widehat{U}$. 
We have argued however that 
the subset $\widehat{S}, \widehat{T}, W,Y$ gives
an appropriate parametrization of any ``universal'' new physics when there is a mass gap.
Note that, if there is no sizable gap between $M_W$ and $\Lambda$, then there is no useful expansion in  $q^2$. 
Indeed in the SM itself there is no gap, and this is why the SM contributions
to all form factors (not just $\widehat S, \widehat T, Y, W$) 
are sizable. 
This is also the case  for
the most interesting region of the supersymmetric parameter space, where some of the spartners are lighter than 200 GeV. 
In any case the data can always be summarized as a measurement of the experimental parameters $\varepsilon_{1,2,3}$ which do not make reference 
to any expansion of any form factor.
The experimental data reported in Table~\ref{tab:data} determine $\varepsilon_{1,2,3}$  as
\begin{equation}\label{eq:LEP1}
\begin{array}{l}
\varepsilon_1= +(5.0\pm 1.1)~10^{-3}\\
 \varepsilon_{2} = -(8.8\pm 1.2)~10^{-3}\\
\varepsilon_{3} =+(4.8\pm 1.0)~10^{-3}
\label{eps123v}
\end{array}
\qquad\hbox{with correlation matrix}\qquad
\rho = \pmatrix{1 & 0.66 & 0.88\cr 0.66 & 1 & 0.46\cr 0.88&0.46&1}\, .
\end{equation}
We recall that the mean values $\mu_i$, the errors $\sigma_i$ and the correlation matrix
$\rho_{ij}$ determine the $\chi^2$ as
$$
\chi^2 =\sum_{i,j} (\varepsilon_i - \mu_i) (\sigma^2)^{-1}_{ij}  (\varepsilon_j - \mu_j),\qquad\hbox{where}
\qquad (\sigma^2)_{ij} = \sigma_i \rho_{ij} \sigma_j\ .$$

In general the new physics corrections to the observables in Table~\ref{tab:datalow},
measured at energies much below the $Z$-pole,
are {\it not} a linear combination of the corrections 
to the $\varepsilon$'s. A linear dependence arises only in universal models where the expansion of the vacuum polarization amplitudes in eq.~(\ref{eq:espansione}) can be
truncated at order $q^2$. Otherwise, when the $q^4$-terms are important to describe the new physics effects, one has again to use the form factors in Table~\ref{tab:STUVXYW}. In this case, 
the low-energy effective Lagrangian at tree-level is
\begin{eqnarray}\nonumber
\Lag_{\rm eff} &=& \Lag_{\rm QED} -4 \sqrt{2} G_{\rm F} ( 1 + \widehat{T}) \sum_{i,j} [ \bar{\psi}_i (T_3 - \sW^2 k Q)
\gamma_{\mu} \psi_i] [ \bar{\psi}_j (T_3 - \sW^2 k Q) \gamma^{\mu} \psi_j]+\\
&&
-2 \sqrt{2}  G_{\rm F} [ \bar{\nu}_L \gamma_{\mu} \bar{\ell}_L ] [ \bar{d}_L \gamma^{\mu} \bar{u}_L ] + \hbox{h.c.}\, ,
\end{eqnarray}
where the sum runs over light SM fermion doublets and
\beq k = 1 + \frac{\widehat{S} - \cW^2(\widehat{T}+W) - \sW^2 Y + 2 \sW\cW X}{\cW^2 -\sW^2} .
\eeq
This Lagrangian can be immediately used for computing 
new-physics corrections to the low-energy observables.
To compute their SM values one needs to include also SM higher order effects.
Given the present uncertainties on low-energy observables,
 their sensitivity to $\widehat S,\widehat T,W,Y$ is about one order of magnitude worse than
the sensitivity of the high-energy observables of Table~\ref{tab:data}.


Before considering the LEP2 data, let us briefly comment on the robustness of the EWPT fit.
When compared with the  SM predictions, as shown in Tables~\ref{tab:data},
\ref{tab:datalow}
there are two
apparently anomalous pieces of data: the NuTeV measurement 
of the $\nu_\mu/$nucleon couplings and $A^b_{FB}$.
The NuTeV anomaly disappears if one conservatively includes among the
uncertainties a possible strange momentum asymmetry
or isospin-violation in the nucleon distributions.
Therefore we prefer not to include NuTeV in the global fit.
Note in any case that the NuTeV results have a minor effect on our best-fit regions.
Similarly, leaving out from the fit the $A_{FB}^b$ asymmetry, whose consistency with the SM (or with any ``universal'' model)  is borderline,  does not modify the determination of the $\varepsilon_i$ in a significant way.\footnote{If one omits these two apparently anomalous pieces of data
the SM gives an excellent fit, with
a best-fit Higgs mass $1.0\sigma$ below its direct limit.
Our global fits includes all data except NuTeV.}


\section{Constraints from LEP2 measurements}\label{LEP2}
For our purposes, the relevant LEP2 observables are the differential cross sections for $e^{+}e^-\to f \bar f$.
Universal new physics modifies them by correcting the transverse part of the $2\times 2$ matrix propagator of the $(Z,\gamma)$ system, which becomes
\beq\label{propagator}
\bordermatrix { & Z&\gamma\cr
Z&G_{ZZ}(s)+\displaystyle\frac{\Delta \varepsilon_1}{s-M_Z^2}-\frac{\varepsilon_{ZZ}}{M_W^2}&G_{Z\gamma}(s)-
\displaystyle\frac{\cW^2(\Delta \varepsilon_1-\Delta
\varepsilon_2)-\sW^2 \Delta \varepsilon_3}{\sW\cW(s-M_Z^2)}-\frac{\varepsilon_{Z\gamma}}{M_W^2}\cr
\gamma&G_{Z\gamma}(s)-\displaystyle\frac{\cW^2(\Delta \varepsilon_1-\Delta
\varepsilon_2)-\sW^2 \Delta \varepsilon_3}{\sW\cW(s-M_Z^2)}-\frac{\varepsilon_{Z\gamma}}{M_W^2}& G_{\gamma\gamma}(s) -\displaystyle
\frac{\varepsilon_{\gamma\gamma}}{M_W^2}}\, ,
\eeq
where $M_Z$ is the pole $Z$ mass and
$G_{ZZ}$, $G_{Z\gamma}$ and $G_{\gamma\gamma}$
are the SM propagators to 1-loop accuracy in a given scheme 
(at tree level $G_{ZZ} = 1/(s-M_Z^2)$, 
$G_{\gamma\gamma} = 1/s$ and $G_{Z\gamma}=0$).
$\Delta \varepsilon_{1,2,3}$ represent the new physics form factor contributions to $\varepsilon_{1,2,3}$ which we discussed in the
 previous section. 
 Finally, 
 $\varepsilon_{ZZ}, \varepsilon_{Z\gamma}$ and $\varepsilon_{\gamma\gamma}$
 are three new observables, measured by LEP2. 
 Note that, since they do not depend on $s$,
 they are equivalent to a specific set of four-fermion operators.
 $\varepsilon_{ZZ}, \varepsilon_{Z\gamma}$ and $\varepsilon_{\gamma\gamma}$
 are induced only by the higher-order new-physics form factors in Table~\ref{tab:STUVXYW} as
\begin{eqnsystem}{sys:vattelapesca}
{\varepsilon_{ZZ}} &=& \cW^2 W -2 \sW\cW X + \sW^2 Y\, ,\\
\varepsilon_{\gamma\gamma} &=&\sW^2 W + 2\sW\cW X + \cW^2 Y\, ,\\
\varepsilon_{Z\gamma} &=& (\cW^2-\sW^2)X+\sW\cW(W-Y)\, .
\end{eqnsystem}
The 1-loop corrected SM propagators will combine with the vertex and box corrections to give the physical SM amplitude.  
Concerning the new physics contributions, notice that the $\Delta \varepsilon_{i}$
have been measured at the per-mille level by EWPT, and agree with the SM.
Moreover at the highest LEP2 energies of $189-207$ GeV these contributions are further suppressed
with respect to the contact terms by a factor  $M_Z^2/s\sim1/4$. 
Therefore, given the LEP2 accuracy of $\sim 1 \%$, 
we can neglect  $\Delta \varepsilon_i$ in eq.~(\ref{propagator})
and directly present
the LEP2 constraints as measurements of  $X,W,Y$.

\medskip

For our purposes the main LEP2 data are the $e\bar{e}\to e\bar{e}, \mu\bar{\mu}, \tau\bar{\tau}, \sum_q q\bar{q}$
cross sections at $\sqrt{q^2} \approx 189,192,196,200,202,205,207\,\GeV$ \cite{LEPEWWG,LEP2}.
Note that the 3 observables $\varepsilon_{ZZ},\varepsilon_{Z\gamma},\varepsilon_{\gamma\gamma}$ can be disentangled through the forward/backward asymmetries, since the  initial state contains both $e_L$ and $e_R$,
which have different $Z$ couplings. 
In the approximation we have described, these data are therefore turned into a direct constraint on $X, Y, W$
\begin{equation}
\begin{array}{rl}
X = \!\!\!&(-2.3\pm 3.5)\,10^{-3}\\
Y =\!\!\! &(+4.2\pm 4.9)\,10^{-3}\\
W = \!\!\!&(-2.7\pm 2.0)\,10^{-3}
\end{array}
\qquad\hbox{with correlation matrix}\qquad
\rho = \pmatrix{1 & -0.96 & +0.84\cr -0.96 & 1 & -0.92\cr +0.84&-0.92&1}\, .
\end{equation}
The error on $X,Y,W$ is at a few per-mille level because the contact terms are enhanced with respect to the SM
amplitude by a factor $s/M_W^2$.
The determinations of the form factors does not improve in a significant way by
including {\sc Hera} and {\sc TeVatron}  data.

\section{Global constraints on $\widehat S$, $\widehat T$, $Y$ and $W$}
\label{fit}

%


\begin{table}
$$ \begin{array}{c|cccc}
\hbox{Type of fit} &10^3  \widehat S &  10^3\widehat T &10^3 Y & 10^3W \\ \hline
\hbox{One-by-one (light Higgs)} & \phantom{-}0.0\pm0.5& 0.1\pm0.6& 0.0\pm0.6& -0.3\pm0.6 \\
\hbox{One-by-one (heavy Higgs)} & \hbox{---} & 2.7\pm0.6&\hbox{---}&\hbox{---}\\ \hline
\hbox{All together (light Higgs)}& \phantom{-}0.0\pm 1.3 & 0.1 \pm 0.9 & 0.1 \pm 1.2 & -0.4 \pm 0.8 \\
\hbox{All together (heavy Higgs)} & -0.9\pm 1.3 & 2.0 \pm 1.0&0.0 \pm 1.2 & -0.2\pm 0.8 \\
\end{array}$$
\caption{\label{tab:fit}\em  Global fit (excluding NuTeV) of dominant form factors
including them one-by-one or all together, with a light ($m_h=115 \GeV$) and with a heavy ($m_h=800 \GeV$) Higgs.}
\end{table}

\begin{figure}[t]
$$\includegraphics[width=18cm]{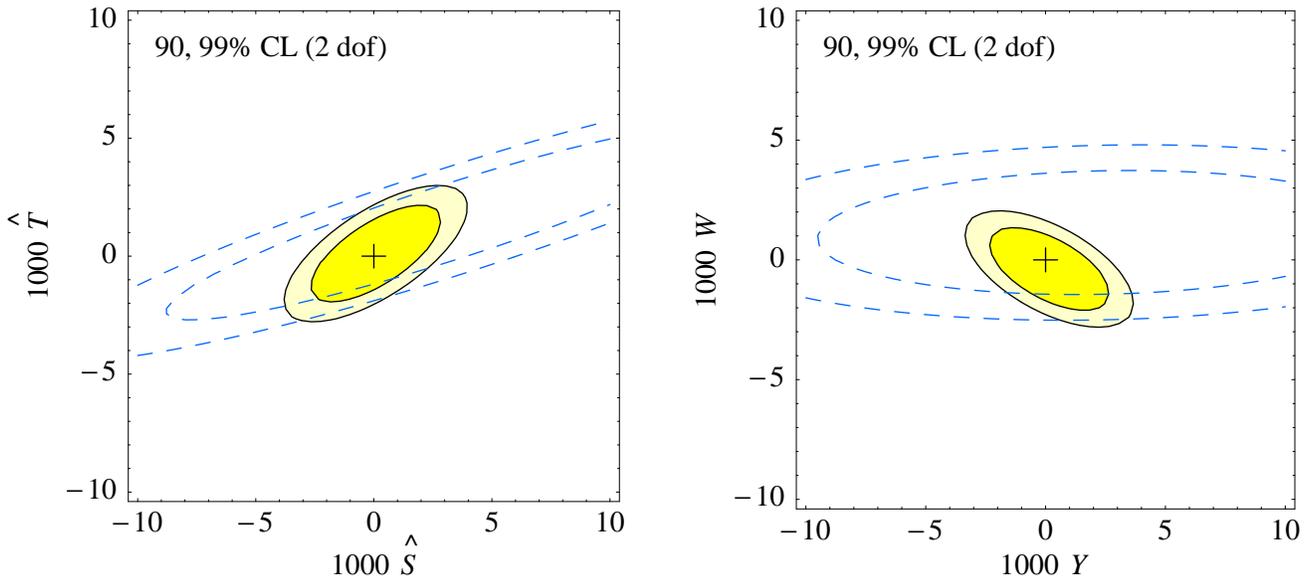}$$
\caption{\label{fig:STWY}\em Allowed values at $90,99\%$ C.L.\
of $(\widehat{S},\widehat{T})$ (for generic $W,Y$)
and of
$(W,Y)$ (for generic $\widehat{S},\widehat{T}$) with $m_h = 115$ GeV.
The dashed lines show the weaker constraints obtained by  the EWPT alone.
}
\end{figure}
\begin{figure}[t]
$$\includegraphics[width=8cm]{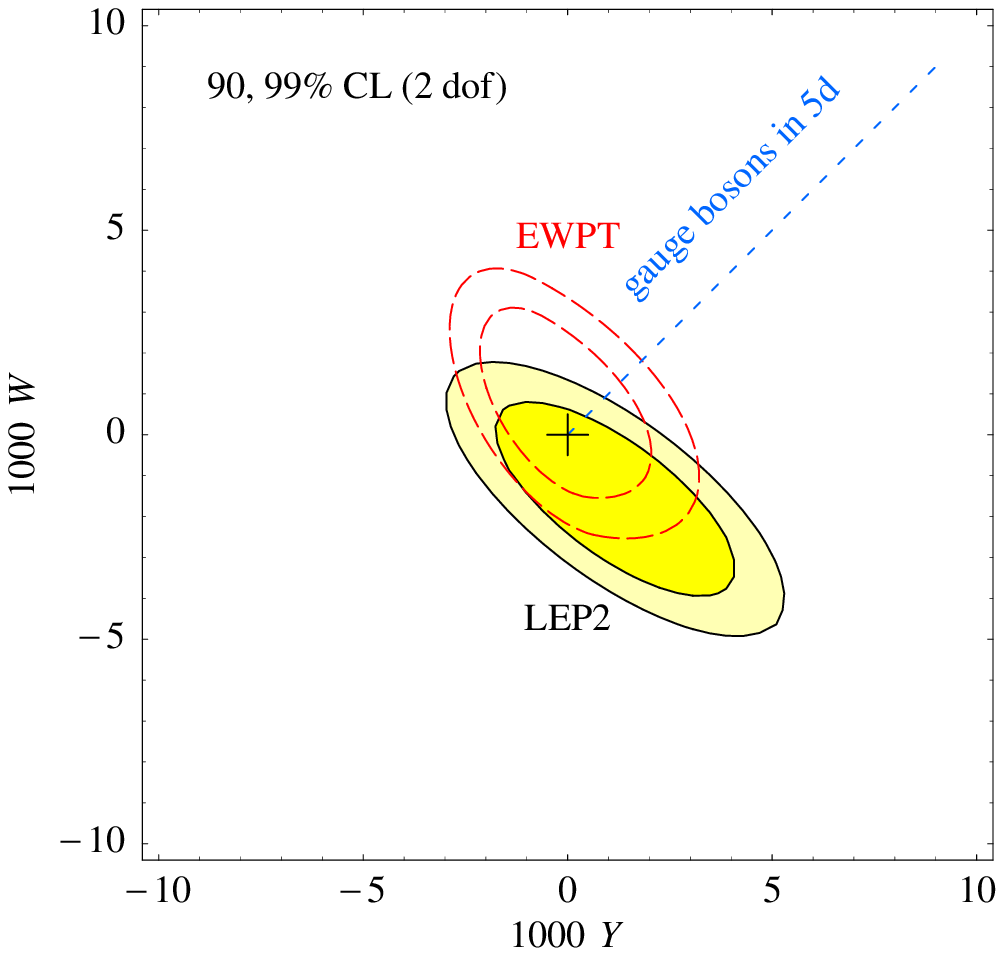}\qquad\includegraphics[width=8cm]{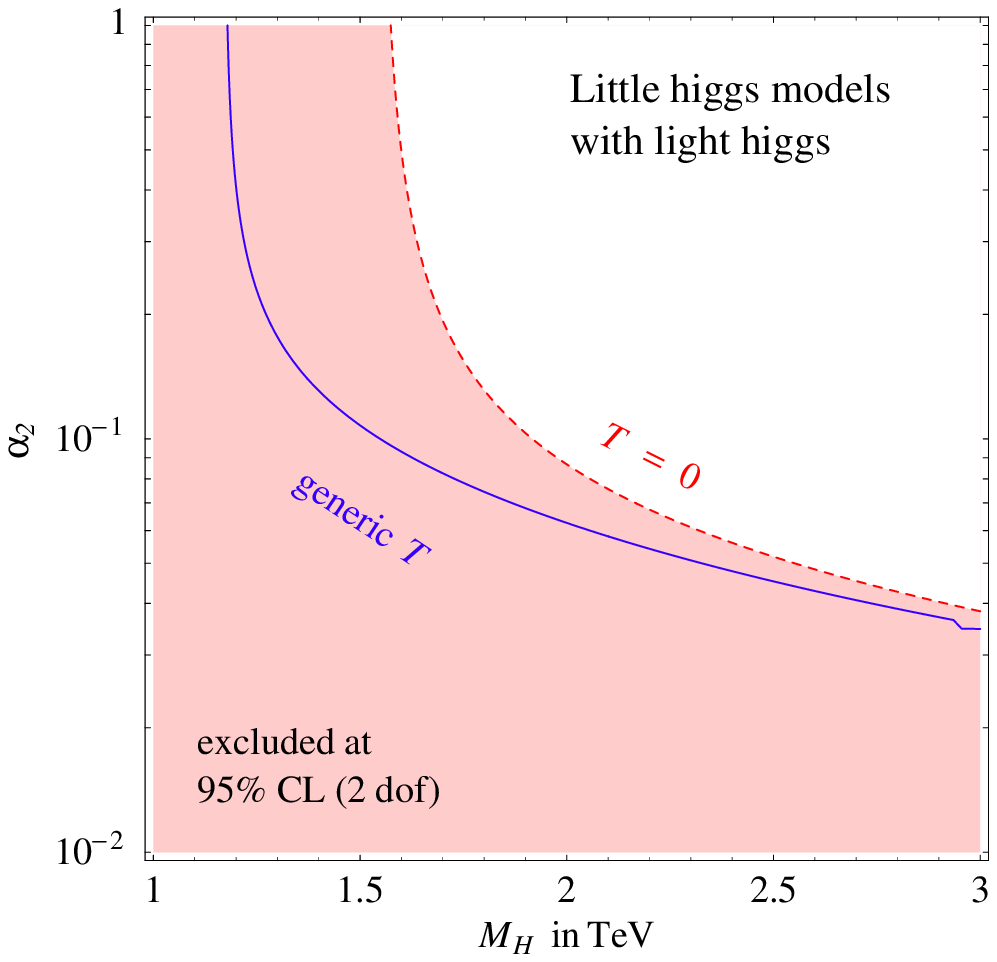}$$
\parbox{8cm}{\caption{\label{fig:WWBB}\em Constraints on the form factors $Y$ and $W$
in models where these are the only new physics effects.
We separately show the impact of EWPT and of LEP2.}}\hspace{1cm}\hfill
\parbox{8cm}{\caption{\label{fig:LittleHiggs}\em Allowed parameters space
of little Higgs models. $M_H$ is the mass of the heavy extra gauge boson,
and $\alpha_2$ the fine structure constant of the extra gauge group.}}
\end{figure}

Adding the LEP2 data to the EWPT
allows
to determine the 4 new-physics form factors $\widehat{S},  \widehat{T}$, $Y$ and $W$.
 The global analysis shows that {\em in a generic ``universal'' model, 
no matter what the Higgs mass is, 
$\hat{S}$, $\hat{T}$, $W$ and $Y$ must be small, at the $10^{-3}$ level}.  
The result of the combined fit (`all together') is shown in Table~\ref{tab:fit}, where we also give the result obtained by adding a single form factor at a time (`one-by-one'), 
both with a light ($m_h=115 \GeV$) and with a heavy ($m_h=800 \GeV$) Higgs. 
The minimum $\chi^2$, relative to the one of the pure SM fit with a light Higgs, does not change significantly in all the cases listed. It would, on the contrary, greatly increase in correspondence with the entries with a blank in Table~\ref{tab:fit}: 
a heavy Higgs can only be compensated by a positive $\widehat{T}$.
A negative $\widehat{S}$ can also allow a satisfactory fit, 
if NuTeV data are included in the global fit.
The correlation matrix relative to the global fits in the last two rows of Table~\ref{tab:fit}, regardless of the Higgs mass, is
\begin{equation}\label{eq:LEP12}
\rho = \pmatrix{1 & 0.68 & 0.65&-0.12\cr 
0.68 & 1 & 0.11 & 0.19\cr 
0.65&0.11&1&-0.59 \cr 
-0.12&0.19&-0.59&1}.
\end{equation}
Some of these correlations are shown in Fig.\fig{STWY}, where we also give the 
allowed regions  that would be obtained from the EWPT of Table~\ref{tab:data}, \ref{tab:datalow} alone.
Such regions are very elongated ellipses because
the precision observables in Table~\ref{tab:data}
are not affected by the following combinations of effects:
 $W=0$, $\widehat{S}=Y=\widehat{T}\cW^2/\sW^2$.
The degeneracy along this direction is only resolved by
 the low energy data of Table~\ref{tab:datalow},
 which however have large uncertainties.
Fig.\fig{WWBB} shows that LEP2 data, beyond resolving this   degeneracy,
provide extra constraints
which are also  competitive with EWPT.
Here we assume that only $W$ and $Y$ are non-vanishing, a physically relevant case (see below), and we show
how the EWPT and LEP2 data separately constrain them.
Whenever $W$ or $Y$ play an important r{\^o}le,
LEP2 data should therefore be taken into account.

\section{Examples of  ``universal'' theories of EWSB and their predictions}
\label{examples}
In this section we  present examples of ``universal'' theories of EWSB
and their predictions for $\widehat S,\widehat T,W$ and $Y$.
As we already explained, 
these are theories in which all new physics effects are contained in the gauge boson
vacuum polarization  amplitudes.
Equivalently, this corresponds to the situation in which the only
interactions of the SM fermions (in addition to Yukawa couplings) 
is via the ${\rm SU}(2)_L\otimes {\rm U}(1)_Y$ currents in eq.~(\ref{fermions}).
We stress that ${\bar W}^a$ and ${\bar B}$ in general are not mass eigenstates corresponding to the
electroweak  gauge bosons.
For instance, in the prototypical little Higgs model of~\cite{Perelstein}, 
based on  the gauge symmetry ${\rm SU}(2)_1\times {\rm SU}(2)_2\times {\rm U}(1)_Y$, the SM fermions
are charged under ${\rm SU}(2)_1$, but the light vectors live in the ``vector'' diagonal subgroup ${\rm SU}(2)_L$ to which
${\rm SU}(2)_1\times {\rm SU}(2)_2$ is broken at the TeV scale. 
Other typical examples are extra-dimensional models  with the SM fermions
confined on a boundary, of which little Higgs theories are often a ``deconstructed'' version.  
 Although the $\bar W$ and the $\bar B$ in eq.~(\ref{fermions})
 are not mass eigenstates,
when studying physics at the electroweak scale they are perfectly good interpolating fields for the light vector bosons. 
By this we mean that the matrix element $\langle W|\bar W|0\rangle$ between the vacuum and the standard bosons is non-zero. Indeed when working at the electroweak
scale there is no need to accurately diagonalize the full mass matrix and find all the eigenvectors, be this a finite or an
 infinite dimensional (Kaluza-Klein) problem. Instead it is often more efficient to find a convenient set of interpolating fields
for the light states and integrate out all the others. 
 It should be stressed that the fields we integrate out are also not exact mass eigenstates in general,
as they  mix  with the chosen interpolating fields. But this does not matter as long as the mass matrix reduced
to the fields we integrate out is non singular. When fermions couple to vector bosons like in eq.~(\ref{fermions}), taking $\bar W,\bar B$ as the low energy fields is the most convenient choice. With this choice, new
physics effects are fully parametrized by vector boson vacuum polarizations.  Using the freedom of choosing the appropriate fields one can drastically simplify the computations
and focus directly on the relevant quantities. For example one immediately sees the equivalence of the 4-fermion interactions mediated by heavy gauge bosons with a suitable ``universal'' effect.

\subsection{Gauge bosons in 5 dimensions}\label{5d}

As a first example we will
consider a model where the SM gauge bosons
 propagate in a flat extra dimension
assumed to be a  $S^1/Z_2$ orbifold of length $L=\pi R$ ($0\leq y \leq L$).
The SM fermions and the Higgs
are assumed to be confined on the same  4 dimensional boundary, 
say, at $y=0$. 

\medskip

Previous analyses 
obtained the following low-energy effective Lagrangian that
describes how heavy KK excitations affect
the low-energy interactions of the SM fields:
\begin{equation}\label{eq:JJ}
\Lag_{\rm eff}=\Lag_{\rm SM} - R^2 \frac{\pi^2}{6}( J^a_\mu J^a_\mu +  J^B_\mu J^B_\mu + J^G_\mu J^G_\mu)+{\cal O}(R^4)\, ,
\end{equation}
where $J$ are the matter currents (fermions plus Higgs) of the three gauge factors of the SM gauge group, normalized as in~\cite{KK}.
The various observables can then be computed by combining
corrections to gauge boson propagators, to their vertices and to 
four-fermion operators.
By appropriately using the tree-level equations of motion, it was recognized in~\cite{KK}
that these corrections are ``universal'' and can be alternatively described by adding
the ``universal'' ${\cal O}_{WW}$ and ${\cal O}_{BB}$ operators of eq.\eq{NRO} with coefficients
$$ c_{BB} = c_{WW} =- (v R)^2 \frac{\pi^2}{6},\qquad c_{WB} = c_H = 0.$$
Therefore $\widehat{S} = \widehat{T}=0$ whereas 
\begin{equation}
W=Y = \frac{(g v\pi R)^2}{6}\, .
\label{all}
\end{equation}
These models provide  a concrete example of a source of new physics
which affects only the higher-order form factors of Table~\ref{tab:STUVXYW}.

\medskip

It is pedagogically useful to see how the same result can be obtained directly, using
the techniques we outlined before.
In this case the r{\^o}le of the interpolating fields
is played by the boundary value of the 5D vectors: ${\bar W}^a_\mu(x)\equiv W^a_\mu(x,y=0)$,
${\bar B}_\mu(x)\equiv B_\mu(x,y=0)$. The heavy fields are given by the field variables at all other points:
$V_\mu^{\rm heavy}(x)=V_\mu(x, y\not=0)$. At tree level the procedure of integrating out the heavy vectors
coincides with solving the 5D equations of motion, while keeping fixed the value of the field at the $y=0$ boundary,
{\it i.e.} keeping the interpolating fields fixed. The low-energy effective action is just the bulk action calculated
on this solution, plus any addition term that may be present at the $y=0$ boundary. The extra dimensional physics
manifests itself only via the vector boson Lagrangian. 
 This procedure  is common in phenomenological applications of the AdS/CFT correspondence
 and was also applied
to Higgsless theories in~\cite{BPR}.
 Perhaps it helps, in order to get a more concrete picture for our separation between bulk
and boundary fields, to consider a discretized (or deconstructed) version of this system \cite{Arkani-Hamed:2001ca}. In this approach the  field
at the boundary is just the gauge field of one group factor in the chain $G_{\rm SM}\otimes G_{\rm SM}\otimes\dots\otimes G_{\rm SM}$
and the bulk fields are the gauge fields of all the other factors.

Applying the above procedure to this model, and confining ourselves to the electroweak sector, we obtain
\begin{equation}
\Lag_{\rm eff}= -\frac{1}{2}\bar W^a_\mu \Pi_{WW}(q) \bar W^{a\, \mu}- 
\frac{1}{2}\bar B_\mu \Pi_{BB}(q) \bar B^\mu\,+\Lag_0\, ,
\label{eff}
\end{equation}
where 
\begin{equation}
\Pi_{WW}(q)=M_L q\tan(qL)\ ,\qquad
\Pi_{BB}(q)=M_B q\tan(qL)\, ,
\label{sigmas}
\end{equation}
are the transverse part of the self-energy contributions from integrating out the bulk
and $\Lag_0$ is the original boundary Lagrangian, involving
the fermions and the Higgs field and possibly extra contributions to the gauge kinetic terms.
$M_{L,B}$ is the inverse squared of the 5D gauge coupling of $W^a$
and $B$ respectively. Notice that, in absence of extra contributions from $\Lag_0$, the $\Pi^{-1}$ are just the boundary to boundary
propagators, with KK poles at $q^2=n^2/R^2$ with $n$ integer.
From eqs.~(\ref{eff}) and (\ref{sigmas})
we  obtain the predictions of the model for 
$\widehat S$,$\widehat T$, $W$ and $Y$.
Since  the bulk is   ${\rm SU}(2)_L\otimes{\rm U}(1)_Y$ invariant,
only $W$ and $Y$ are nonzero:
\begin{equation}
 W=\frac{g^2}{2}M^2_W\Pi_{WW}^{\prime\prime}(0)
=\frac{g^2}{3} M^2_WM_LL^3\ ,\qquad
 Y=\frac{g^{\prime 2}}{2}
M^2_W\Pi_{BB}^{\prime\prime}(0)=\frac{g^{\prime\, 2}}{3}
M^2_WM_BL^3\, .
\end{equation}
If no boundary kinetic terms are present in the theory we have 
\begin{equation}
M_LL=1/g^2\ ,\qquad
M_BL=1/g^{\prime\, 2}\, ,
\label{gare}
\end{equation}
and the model consists of  only one parameter  $R$ with $W$ and $Y$ as in eq.~(\ref{all}).

\bigskip

Fitting the  latest electroweak data in Tables~\ref{tab:data},  \ref{tab:datalow}  gives the constraint
 $1/R > 4.5\TeV$ at 95\% CL.
If $1/R$ is close to its lower bound, the Higgs can be somewhat heavier than
what allowed by a pure SM fit, because
5D bulk effects  partially compensate the effects of a heavy Higgs~\cite{newref}.
The bound from LEP2 alone is $1/R> 6.3\TeV$ at 95\% CL, 
and does not depend on the Higgs mass.
The combined bound is $1/R>6.4\TeV$ at 95\% CL (see also~\cite{landsberg}):
the Higgs can no longer be heavier than what allowed by a pure SM fit.
This can be also seen from Fig.\fig{WWBB}, where the diagonal dashed line
corresponds to the parameter space of this model.
Related models where only the SU(2)$_L$ or only the U(1)$_Y$ gauge bosons live in the 5th dimension,
would be represented by vertical and horizontal lines, respectively.

As an aside remark, we note that similar considerations apply to the QCD sector as well.
Using the same technique we can parametrize the leading effect of the gluon KK modes
by  a $q^4$ correction to the gluon self-energy, $Z$, rather than by the effective 4-fermion operator $J_\mu^G J_\mu^G$.
In the absence of boundary kinetic terms we have $Z=W=Y$, see eq.~(\ref{all}).
We estimate that LHC, in absence of more striking phenomena, should test the $Z,W,Y$ 
form factors with a precision of few $10^{-3}$.

\medskip

In the rest of this paper we will analyze less simple ``universal'' models:
in order to obtain simple and correct results it now becomes really important
to recognize them as ``universal''.

\subsection{Gauge bosons and Higgs in 5 dimensions}
Assuming that  the Higgs, instead of being confined on the boundary,
 also propagates in the 5th dimension,
  the 5D bulk breaks both custodial and isospin symmetries so that all
the 4 parameters are generated: 
\begin{equation}
\widehat S=\frac{2}{3} M^2_WL^2\ ,\qquad
\widehat T=\frac{1}{3g^2}\frac{M^2_WL}{M_B}\ ,\qquad
W=\frac{g^2}{3}M^2_WM_LL^3\ ,\qquad
 Y=\frac{g^{\prime\, 2}}{3}M^2_WM_BL^3\, .
\label{hbo}
\end{equation}
Notice in particular that $\widehat T$ comes out proportional to the 5D hypercharge coupling $1/M_B$.
In the absence of boundary kinetic terms (for both the the gauge bosons and the Higgs)
the use of eq.~(\ref{gare}) shows that all the contributions in eq.~(\ref{hbo})
are comparable.
In this case, the 95\% C.L.\ limit from EWPT alone now gives $1/R>3.8\TeV$, 
while the limit from
LEP2 alone remains, as in the previous case, $1/R>6.3\TeV$.
The combined limit is $1/R>6.1\TeV$ at $95\%$ C.L.
The upper limit on the Higgs mass negligibly varies with respect to the pure SM case.

\subsection{Little Higgs models}

Little Higgs models are based on  ``deconstructed''
 extra dimensional models in which the Higgs 
arises as a pseudo-Goldstone boson (PGB).
These models  contain products  of the same gauge group, e.g.
${\rm SU(2)}_1\otimes {\rm SU(2)}_2 \otimes\cdots$, 
with  the SM fermions (usually)  charged only under one of them.
According to the previous discussion,
also these theories  can  be  categorized as ``universal''.
As an illustrative example 
we will focus on the
${\rm SU(2)}_1\otimes {\rm SU(2)}_2\otimes {\rm U(1)}_Y$
 model discussed in~\cite{Perelstein},
based on  the  previous little Higgs models of~\cite{littleH}.
Without including LEP2 data, the  EWPT  analysis
for this model was  already carried out in~\cite{Perelstein}.
To illustrate here the simplicity of  our procedure we present the same analysis, 
including at the same time  the LEP2 data.

The model has a global symmetry SU(5) of which
 only a subgroup  ${\rm SU(2)}_1\otimes {\rm SU(2)}_2\otimes {\rm U(1)}_Y$
is gauged.
By imposing the symmetry breaking pattern
SU(5)$\rightarrow{\rm SO}(5)$, an SU(2) doublet
of PGB appears in the spectrum corresponding to the 
SM Higgs. Fermions are only charged under  ${\rm SU}(2)_1 \otimes {\rm U}(1)_Y$. 
Integrating out the other vectors we get the effective Lagrangian
for the light fields
\bea\label{efflan}
\Lag&=&-\frac{1}{4g^2_1}  \bW ^a_{\mu\nu}\bW^{a\, \mu\nu}
-\frac{1}{4g^{\prime 2}}  \bB_{\mu\nu} \bB^{\mu\nu}
+\frac{f^2}{4}  \bar W^a_\mu \bar W^{a\mu} 
+\frac{(1-c)f^2}{4}\bB_\mu \bB^{\mu}
+\frac{f^2\zeta}{4}\bW^3_\mu \bW^{3\mu}\\
&&- \frac{f^2}{4}(1-c)\bW^3_\mu \bB^{\mu}
+\frac{g^2_2}{16}\frac{(1+c)^2f^4}{q^2-{g^2_2 f^2}/{2}}
\bW^+_\mu \bW^{-\mu}+
 \frac{g^2_2}{32}\frac{ [ (1+c+2\zeta) \bW^3_\mu
+  (1-c) \bB_\mu]^2  f^4}{q^2-{g^2_2 f^2(1+\zeta)}/{2}}\, ,\ \ \ \ \ 
 \nonumber
\eea
where $g_1$ and $g_2$ are the gauge couplings of SU(2)$_1$ and SU(2)$_2$,
$f$ is the scale at which they are broken into the usual SU(2)$_L$
with coupling $1/g^2 = 1/g_1^2 + 1/g_2^2$,
$c=\cos(\sqrt{2}v/f)$, $\zeta = \sin^4(v/\sqrt{2}f)/2$, $\tan\psi=g_1/g_2$. 
From  eq.~(\ref{efflan}) we  extract, at dominant order in $v/f$,
\beq
\widehat S=\frac{\sin^2\psi}{g^2}\frac{ M^2_W}{f^2},\qquad
\widehat T= {\cal O}(\frac{v}{f})^6
,\qquad
W=\frac{2\sin^4\psi}{g^2}\frac{ M^2_W}{f^2},\qquad
Y={\cal O}(\frac{v}{f})^6\, .
\label{lh}
\eeq
We see that for for $g_1 \sim g_2 \sim g$ both $S$ and $W$ are relevant,
while for large $g_2$
the dominant effect appears in $\widehat S$
and only depends on the mass
of the heavy charged boson,
$M_H^2  = (g_1^2 + g_2^2) f^2/2$.
The model also involves an isospin triplet PGB, which can generate a potentially relevant, but model dependent,
contribution to $\widehat T$.
Here we  consider both the case of a small $\widehat T$ and the case
of arbitrary $\widehat T$. 
The allowed parameter space is shown in Fig.~\ref{fig:LittleHiggs} for both cases.
For appropriate values of $\widehat T$ it is possible to fit data with
a Higgs mass heavier than what allowed by a pure SM fit, while
this is not possible for small $\widehat T$.

\subsection{Higgsless models}\label{higgsless}

It is well known \cite{tech,pt} that  in the traditional Higgsless scenario, technicolor, $ T$ and $ S$
are the only relevant parameters. 
However one can imagine a more general situation where also $W$ and $Y$ are relevant,
signifying that the gauge bosons themselves are ``composite'' at the TeV scale. 
This situation can be effectively realized in regions of the parameter
space of the recently proposed 5-dimensional Higgsless theories. 
We will now focus on a representative of this class of models \cite{BPR}.
Our conclusions are however very general and apply with minor 
modifications to all the other models, in particular to those on warped spaces
(see~\cite{Csaki:2003dt,higgslessrefs,Cacciapaglia,dhlr}).
Our model \cite{BPR} is based on the gauge group 
${\rm SU}(2)_L\otimes {\rm SU(2)}_R\otimes {\rm U}(1)_{B-L}$
with a flat compact extra dimension of length $L=\pi R$. 
The condition at the boundary (the SM-boundary)
where the fermions are mostly 
localized breaks the gauge group down to the SM group,
whereas at the other boundary (the EWSB-boundary)
the preserved symmetry   
is ${\rm SU(2)}_{L+R} \otimes {\rm U(1)}_{B-L}$ \cite{Csaki:2003dt, adms}.
The advantage of this set up is that both the 5D bulk 
and the EWSB-boundary  respect a custodial  symmetry. 

After integrating out the 5D bulk at tree level,
the effective Lagrangian for the fields at the  SM-boundary is
\bea
\Lag&=&-\frac{1}{4g^2} \bar W^a_{\mu\nu}\bar W^{a\, \mu\nu}
-\frac{1}{4g^{\prime 2}} \bar B_{\mu\nu}\bar B^{\mu\nu}+\nonumber\\
&&- \frac{1}{2}\bar W^a_\mu \Delta\Pi_{WW}(q) \bar W^{a\mu} 
-\bar W^3_\mu  \Delta\Pi_{WB}(q)\bar B^\mu-\frac{1}{2}
\bar B_\mu  \Delta\Pi_{BB}(q)\bar B^\mu\, .
\label{himo}
\eea
The vacuum polarizations $ \Delta\Pi_V$ of eq.~(\ref{himo})
are related to  the corresponding 
$\Sigma_V$ calculated in~\cite{BPR} as
$ \Delta\Pi_{WW}(q)=2 \Sigma_{WW}(iq)$,
$ \Delta\Pi_{BB}(q)=2 \Sigma_{BB}(iq)$,
$ \Delta\Pi_{WB}(q)= \Sigma_{WB}(iq)$.
Moreover in order to keep the same normalization of~\cite{BPR} we work on the double covering of
the orbifold.
The $ \Delta\Pi$'s depend on the 5D gauge coupling $M_{L,R,B}$ and on the 
kinetic  coefficients $Z_{W,B}$ localized at the EWSB-boundary.
At leading order,  $M_W$ is given by
\beq
M_W^2=\frac{2g^2 M_LM_R}{(M_L+M_R)L}\equiv  2g^2 \frac{M}{L}\, .
\eeq
Working in the limit
$R M_W\ll 1$ (otherwise there would be extra light states),
we can expand in $q^2$ the  $\Pi(q)$'s and   calculate 
the contribution to the $\widehat S,...,Y$ parameters
\begin{eqnsystem}{sys:higgsless}
\widehat S&=&
g^2
\frac{4}{3}M L\left [1+\frac{3}{4}z_W\right]\, ,\label{S}\\
\widehat T&=&\widehat U=0\, ,\label{T}\\
X&=&
gg^\prime M_W^2ML^3
\left[\frac{28}{45}+z_W+\frac{1}{2}z_W^2\right]\, ,\\
W&=&
g^2\frac{M_W^2ML^3}{1-y_L}\left[
\frac{28}{45}y_L+ \frac{2}{45}+z_W y_L+
\frac{1}{2}z_W^2y_L\right]\label{W}\, ,\\
Y&=&
g^{\prime\, 2}\frac{M_W^2ML^3}{1-y_R}\left[
\frac{28}{45}y_R+ \frac{2}{45}+z_W y_R+
\frac{1}{2}z_W^2y_R\right]\nonumber\\
&&+g^{\prime\, 2}M_W^2M_BL^3\left[
 \frac{2}{3}+z_B+
 \frac{1}{2}{z_B^2}\right]\, ,\label{Y}
\end{eqnsystem}
where we have defined 
\bea
&&y_L\equiv \frac{M_L}{M_L+M_R}\ ,\quad\quad y_R\equiv \frac{M_R}{M_L+M_R}=1-y_L\, ,\\
&&z_W\equiv\frac{Z_W}{(M_L+M_R)L}\ ,\quad\quad z_B\equiv\frac{Z_B}{ M_BL}\, .
\eea
For warped backgrounds,  only  the numerical
coefficients are slightly modified (see~\cite{BPR}
for the contribution to $\widehat S$ in different backgrounds).

\medskip

In~\cite{BPR} the limit 
\beq
M_LL,M_RL, M_BL,Z_{W,B}\ll 1/g^2\, ,
\label{4d}
\eeq
was taken.
In this case, only the contribution to $\widehat S$ is relevant.
Notice that $\widehat S$ grows with the inverse 5D coupling $M$. More precisely parametrizing the 5D
loop expansion parameter as\footnote{By applying naive dimensional analysis~\cite{nda2} the scale at which
the 5D theory becomes strongly coupled is $48\pi^3 M$. 
When $\ell_5\sim 1$ the theory is strongly coupled already
at the energy of the lightest KK mode, 
so that the 5D description is never valid, and predictivity is totally lost
\cite{unita}.
Notice that our definition of $\Lambda$ 
differs by a factor 2 with respect to~\cite{BPR}. This is to
account for the proper normalization of the path integral when working over the doubly covered orbifold, though we are
aware that NDA works only within factors of order 1. A more detailed analysis of perturbativity in these models will be 
presented in~\cite{papucci}.}
  $\ell_5=1/(48 \pi^3 MR)$ we have 
\begin{equation}
{\widehat S}=\frac{g^2}{36\pi^2}\frac{1}{ \ell_5}\, .
\label{Sloop}
\end{equation}
For $\ell_5 \ll 1$ (necessary to have a reliable 5D gauge coupling expansion),
one has $\varepsilon_3\simeq \widehat{S}\gg 10^{-3}$,
indicating that (marginal) agreement with the data can only be obtained in the region where $\widehat S$ is not calculable.
In this respect a Higgsless theory in 5D does not fare better than a generic strongly coupled and incalculable
4D one. In the warped model of~\cite{Csaki:2003dt}  the 5D coupling is fixed by the 4-dimensional one and by the Planck to Weak
scale ratio in such a way that ${\widehat S}$ is predicted to be $\sim 10^{-2}$~\cite{BPR}, 
which is excluded at many standard deviations.

\begin{figure}[t]
$$\includegraphics[width=8cm]{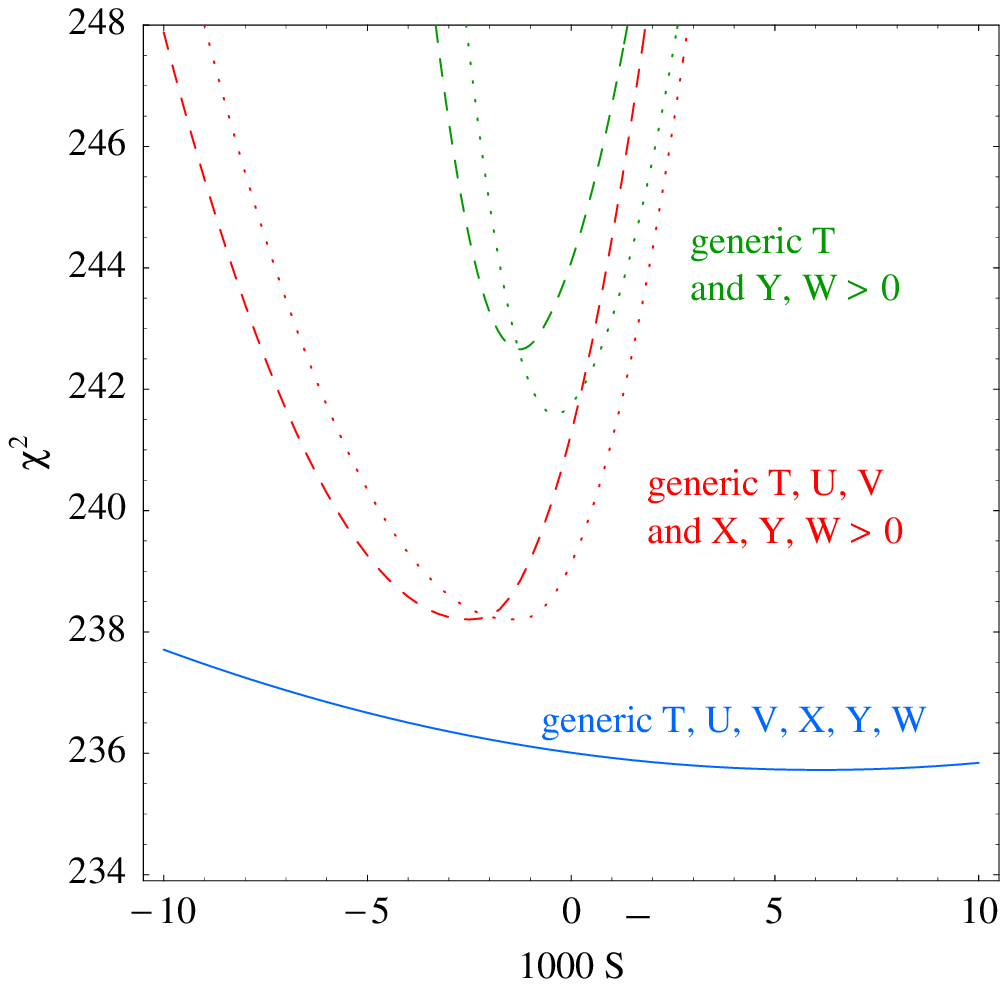}\qquad\includegraphics[width=8cm]{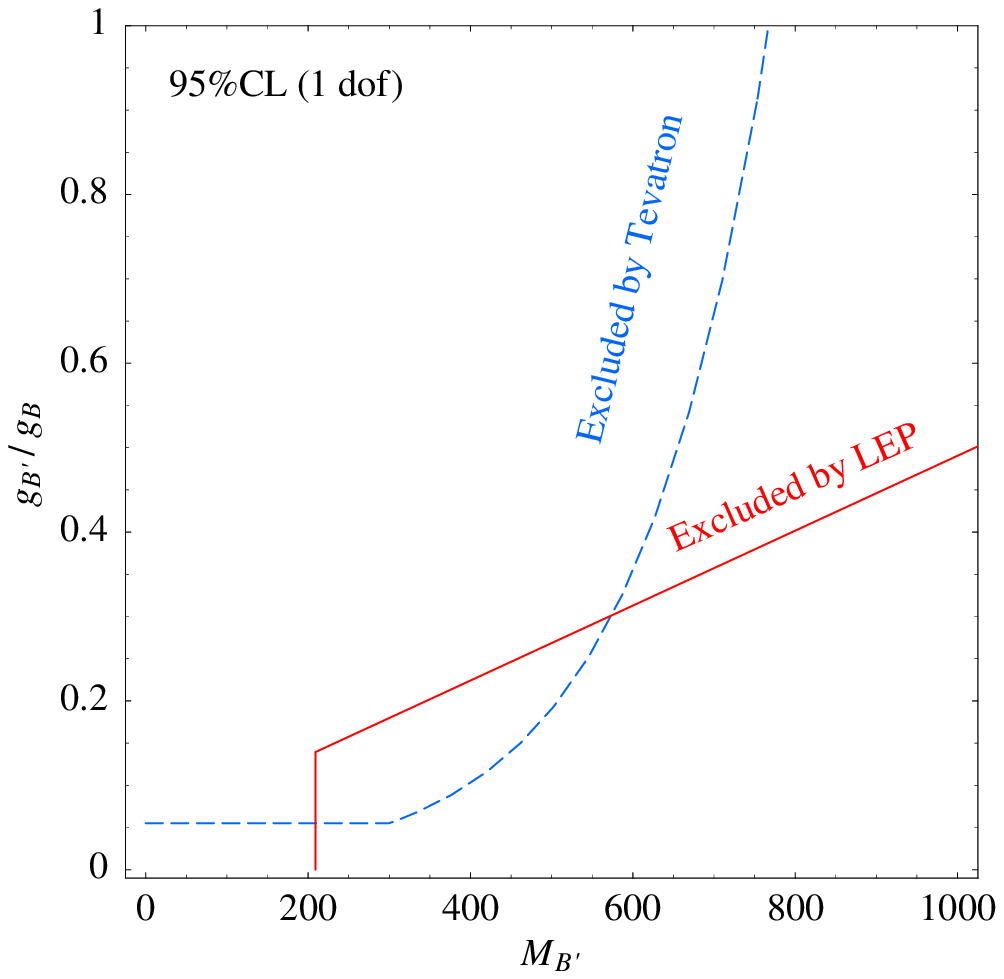}$$
\parbox{8cm}{\caption{\label{fig:Higgsless}\em  $\chi^2(\widehat S)$ for $m_h = 115\GeV$ (dotted), $800\GeV$ (dashed).
The $n$-$\sigma$ range for $\widehat S$ corresponds to $\chi^2 < \chi^2_{\rm min} +n^2$.
Higgsless models that predict $\widehat{S}\approx + \hbox{\rm few} \cdot 10^{-3}$
and $X,Y,W>0$ are excluded.
}}\hspace{1cm}\hfill
\parbox{8cm}{\caption{\label{fig:B}\em Constraints on a massive $Z'$ boson
coupled to hypercharge, as function of its mass (in $\GeV$) and of its gauge coupling
(normalized with respect to the coupling of the SM gauge boson).}}
\end{figure}

One may ask, however,  whether by relaxing eq.~(\ref{4d}), thus extending the model to its full parameter space,  
larger values  of $ \widehat{S}$ 
could be made compatible with the data \cite{Cacciapaglia}. 
For instance, if ${\widehat S}\sim 10^{-2}$ were acceptable,
even at the price of some fine-tuning, then the model would be perturbative and testable in a reasonable range of energies with strikingly new phenomena.
By inspection of eqs.~(\ref{S}-\ref{W}), one can see, as expected,  that only 
$W$ and $Y$ can become sizable  enough to compete with $\widehat S$
and possibly compensate its effects.
In fact, apart from $g,g'$ factors, one has $X\sim \widehat S^2\ll \widehat S$. 
Notice also that one cannot play with $z_W$ as it must
be positive to avoid ghosts. On the other hand, while keeping $\widehat S$ 
and $M_W$ fixed, one can enhance $W$ and $Y$.
In the case of $W$ this occurs for $M_L=M/(1-y_L)\gg M$, while in the case of $Y$ it can occur both with
$M_B\gg M$ or with $z_B\gg 1$. The choices of parameters for which $W,Y$ become relevant correspond to delocalizing 
to the bulk the electroweak gauge bosons. 
 From the point of view of an equivalent purely 4D 
strongly coupled theory this can be interpreted as the gauge bosons being
composites at the electroweak breaking scale $1/R$.
However when this effect is achieved by taking a large $z_B$ the delocalization is associated to the presence of
a new light vector boson localized close to the EWSB-boundary. 
As we will comment below, this possibility is severely
limited by the {\sc TeVatron} data on $Z'$ searches \cite{tevatron}. 
This problem does not arise when $W,Y$ are enhanced by increasing
$M_L$ and $M_B$.
 Notice finally that, throughout the parameter space, $\widehat S$, $W$ and $Y$ 
are all positive. 

We are thus lead to consider a possible fit of the data in terms of positive $\widehat {S}, W, Y$. However, even allowing for the presence of an unknown and possibly large one-loop contribution to the parameter $\widehat T$ from custodial-breaking top effects, this cannot make the model acceptable since the parameter $\widehat S$ has to be small in any case (see Table~\ref{tab:fit}). This remains true even for 
generic $\widehat T$, $\widehat U$, $V$ and positive $X, W, Y$, 
as shown in Fig.~\ref{fig:Higgsless}. We stress that the conclusions implied by Fig.~\ref{fig:Higgsless} apply equally well to the warped Higgsless models
\cite{Csaki:2003dt,higgslessrefs,Cacciapaglia,dhlr}. This is because the expression for $\widehat S$ in terms of the 5D loop expansion parameter 
is qualitatively the same as eq.~(\ref{Sloop}) \cite{BPR}, while $W$ and $Y$ remain positive.

This conclusion is based on the expansion in eq.~(\ref{eq:espansione}) of the self energies.
This expansion in $q^2$ is a very good approximation when the new states have a mass significantly
higher than LEP2 energies. However for our purposes there
is no need to perform a more dedicated analysis even if the new states have rather low mass,
say around $300$ GeV. Indeed to account more accurately for a relatively light resonance of mass $m$
it is enough to correct the contact interactions in eq.~(\ref{propagator}) to include the pole:
$\varepsilon_{\gamma\gamma}\to \varepsilon_{\gamma\gamma} m^2/(m^2-q^2)$ and similarly for the others.
Therefore our approximation slightly underestimates the effects that we want to avoid.
Furthermore, {\sc TeVatron} data~\cite{tevatron}
directly constrain such particles~\cite{dhlr}.
Within the present model, for  large $Z_B/M_BL$
one KK state of the $B$-boson becomes light with a mass $M_{B'}^2\simeq  2M_B/(Z_BL)$.
This particle acts approximately like a heavy hypercharge vector boson. However, its coupling
to matter instead of being $g'$ is equal to 
$g_{B'}\simeq {\sqrt Z_B} g^{\prime\, 2}\ll g'$.
Because of this quadratic dependence in $g'$ this vector looks also very similar to a $\rho$-meson
coupling to electrons via vector meson dominance.
The relation $g_{B'}^2/M_{B'}^2=Y g^{\prime\, 2}/M_W^2$ also holds between the coupling, the mass and the parameter $Y$.
As shown in Fig.\fig{B}, at small mass and small coupling Tevatron constraints are even more
significant than the LEP constraints.

\medskip

In conclusion, 
5d Higgsless models predict a $\widehat{S}$ not compatible with data.
In order to resurrect such models one could add
some ad-hoc new physics that compensates
the too large correction to $\widehat{S}$.

\section{Conclusions}
Alternative models of EWSB keep being proposed, based on different motivations. It is therefore essential to compare them with experiments in a clean and effective way. Most of the time the models proposed are of ``universal'' type, i.e.\ they modify the SM only in the self-energies of the vector bosons. As we have seen, to recognize that a model is ``universal'' may require an optimized definition of the effective vector bosons themselves. This is the case for several models that recently  received attention.

Once this is done, general  ``universal'' models with heavy new physics can be conveniently compared with the low-energy data in terms of four form factors, $\widehat{S}, \widehat{T}, W, Y$. These four parameters are strongly constrained by the combination of the EWPT and of the LEP2 $e\bar{e}\to f\!\bar{f}$
data. In a combined fit with the inclusion of a light Higgs mass, all these form factors are consistent with zero, within $10^{-3}$ uncertainties. 
When all the four form factors are allowed to vary simultaneously, a fit of the data is possible even for a heavy Higgs, with a moderate deviation from zero of $\widehat{S}$ and especially $\widehat{T}$. 
This  relaxation of the usual SM upper bound on the Higgs mass 
however requires some fine-tuning of the new-physics form factors~\cite{NRO}:
in the examples we studied this is the case only in the little Higgs model, 
when $\widehat{T}$ is treated as a free parameter.
It looks to us more important that all the new-physics parameters must remain small in any case, as true in particular for $\widehat{S}$.

This analysis can be conveniently applied to several models of recent interest, like little Higgs models, models with gauge bosons in extra dimensions or Higgsless models in 5D. In this way it is straightforward to see the constraints on their respective parameter space. As an explicit application it is in particular possible to explore Higgsless models of EWSB in their full range of parameters. We find that, when calculable, all the proposed models do
not provide a viable description of electroweak symmetry breaking.

\paragraph{Acknowledgments}
We thank I. De Bonis, R. Tenchini and P. Wells for providing and explaining
us the results on $e\bar{e}\to e\bar{e}$ data. We also thank C.Cs{\'a}ki and C. Grojean for discussions
and T. Gregoire for discussions and for clarifications on little Higgs models.
The work of AP was  supported  in part by 
the MCyT and FEDER Research Project
FPA2002-00748 and DURSI Research Project 2001-SGR-00188.
The work of RB was  supported  in part by MIUR and by the EU under TMR contract
HPRN-CT-2000-00148.

\frenchspacing\footnotesize\begin{multicols}{2}
\end{multicols}
\end{document}